\documentstyle[12pt]{article}
 
\begin{document}


\newcommand{\be}[1]{\begin{equation}\label{#1}}
\newcommand{\beq}{\begin{equation}}
\newcommand{\ee}{\end{equation}}
\newcommand{\beqn}[1]{\begin{eqnarray}\label{#1}}
\newcommand{\eeqn}{\end{eqnarray}}
\newcommand{\bd}{\begin{displaymath}}
\newcommand{\ed}{\end{displaymath}}
\newcommand{\mat}[4]{\left(\begin{array}{cc}{#1}&{#2}\\{#3}&{#4}\end{array}
\right)}
\newcommand{\matr}[9]{\left(\begin{array}{ccc}{#1}&{#2}&{#3}\\
{#4}&{#5}&{#6}\\{#7}&{#8}&{#9}\end{array}\right)}
\def\simlt{\mathrel{\lower2.5pt\vbox{\lineskip=0pt\baselineskip=0pt
           \hbox{$<$}\hbox{$\sim$}}}}
\def\simgt{\mathrel{\lower2.5pt\vbox{\lineskip=0pt\baselineskip=0pt
           \hbox{$>$}\hbox{$\sim$}}}}
\def\unity{{\hbox{1\kern-.8mm l}}}
\def\epr{E^\prime}
\def\al{\alpha}
\def\ga{\gamma}
\def\Ga{\Gamma}
\def\om{\omega}
\def\OM{\Omega}
\def\la{\lambda}
\def\La{\Lambda}
\newcommand{\eps}{\varepsilon}
\def\ep{\epsilon}
\newcommand{\ov}{\overline}
\renewcommand{\to}{\rightarrow}
\def\mcirc{{\stackrel{o}{m}}}
\newcommand{\bx}{\bar{\rm X}} 
\newcommand{\wx}{{\rm X}} 
\newcommand{\bv}{\bar{\rm V}} 
\newcommand{\wv}{{\rm V}} 
\newcommand{\tl}{\tilde{l}} 
\newcommand{\tq}{\tilde{q}} 
\newcommand{\tuc}{\tilde{u}_c} 
\newcommand{\tdc}{\tilde{d}_c} 
\newcommand{\tec}{\tilde{e}_c} 
\newcommand{\TQ}{\tilde{Q}} 
\newcommand{\TU}{\tilde{U}}
\newcommand{\TE}{\tilde{E}} 
\newcommand{\TUC}{\tilde{U}_c} 
\newcommand{\TEC}{\tilde{E}_c} 
\newcommand{\TQC}{\tilde{Q}_c} 
%
%
\makeatletter
\newcounter{alphaequation}[equation]
\def\thealphaequation{\theequation\alph{alphaequation}}
%
\def\eqnsystem#1{
\def\@eqnnum{{\rm (\thealphaequation)}}
\def\@@eqncr{\let\@tempa\relax
\ifcase\@eqcnt \def\@tempa{& & &}
\or \def\@tempa{& &}\or \def\@tempa{&}\fi\@tempa
\if@eqnsw\@eqnnum\refstepcounter{alphaequation}\fi
\global\@eqnswtrue\global\@eqcnt=0\cr}
\refstepcounter{equation}
\let\@currentlabel\theequation
\def\@tempb{#1}
\ifx\@tempb\empty\else\label{#1}\fi
\refstepcounter{alphaequation}
\let\@currentlabel\thealphaequation
\global\@eqnswtrue\global\@eqcnt=0
\tabskip\@centering\let\\=\@eqncr
$$\halign to \displaywidth\bgroup
  \@eqnsel\hskip\@centering
  $\displaystyle\tabskip\z@{##}$&\global\@eqcnt\@ne
  \hskip2\arraycolsep\hfil${##}$\hfil&
  \global\@eqcnt\tw@\hskip2\arraycolsep
  $\displaystyle\tabskip\z@{##}$\hfil
  \tabskip\@centering&\llap{##}\tabskip\z@\cr}

\def\endeqnsystem{\@@eqncr\egroup$$\global\@ignoretrue}
\makeatother





       


\begin{flushright}
hep-ph/9607363 ~~~~~ INFN-FE 03/96 \\ 
July 1996 \\
\end{flushright}
\vspace{10mm}

\begin{center}
{\large \bf PROBLEM OF FLAVOUR IN SUSY GUT \\ 
AND HORIZONTAL SYMMETRY} 
\footnote{Based on talks given at Int. Workshop {\em `SUSY 96'}, 
Univ. of Maryland, 29 May - 1 June 1996 (to appear on Proceedings), 
and II US-Polish Workshop {\em `Physics from Planck Scale to 
Electroweak Scale'}, Warsaw, 28-30 March 1996. } 
\end{center} 

\vspace{0.3cm}
\centerline{\large Zurab Berezhiani }
\vspace{4mm} 
\centerline{\it INFN Sezione di Ferrara, 44100 Ferrara, Italy,}
\baselineskip=12pt
\centerline{\it Institute of Physics, Georgian Academy of Sciences, 
380077 Tbilisi, Georgia} 
\vspace{1.9cm}

\begin{center}
{\bf Abstract} 
\end{center} 
\vspace{2mm}

The concept of non-abelian horizontal symmetry $SU(3)_H$ 
can greatly help in understanding the fermion and sfermion 
flavour structures in supersymmetric grand unification. 
For the sake of demonstration the $SU(5)\times SU(3)_H$ model, 
suggested earlier in ref. \cite{PLB85}, is revisited. 
We show that under very simple and natural assumption it 
links the sfermion mass pattern to those of fermions in a 
remarkable way. 
All dangerous supersymmetric flavour-changing contributions are 
naturally suppressed in a general case, independently of the
concrete texture for fermion mass matrices. 
Nevertheless, within this framework we present 
an example of predictive model for fermion masses and mixing, 
which leads to 7 consistent predictions for the low energy 
observables. 



\newpage
\section{Introduction}


The two most promising ideas beyond the Standard Model, 
supersymmetry (SUSY) and grand unification theory (GUT),
can be united by a simple relation \cite{ICTP}: 
\be{Love}
\mbox{SUSY} + \mbox{GUT} = \mbox{LOVE} 
\ee
In support of this formula one can mention the potential of SUSY 
to solve the personal problems of GUT (gauge hierarchy and 
doublet-triplet splitting), an impressive fact of the MSSM gauge 
couplings unification in minimal $SU(5)$, 
$b-\tau$ Yukawa unification and its impact on top mass, etc. 

There are, however, family (flavour) problems, 
which in the context of supersymmetric theory have two aspects: 
fermion flavour and sfermion flavour. 
First aspect, questioning the origin of family replication, 
quark and lepton mass spectrum and mixing, etc. 
(for a review, see ref. \cite{ICTP}), has no appealing 
solutions within the naive (minimal) LOVE models where the  
Yukawa constants remain arbitrary (hereafter LOVE is defined 
by eq. (\ref{Love})). However, there are 
realistic and predictive LOVE frameworks based on $SO(10)$ 
\cite{so10} or $SU(6)$ \cite{su6} symmetries which provide 
satisfactory explanations to the fermion flavour features. 

Second aspect, which is a specific of SUSY, questions 
the origin and pattern of the soft SUSY breaking (SSB) terms, 
or in other words the mass and mixing spectrum of superpartners. 
There are no direct experimental data on sfermion mass pattern.  
Theoretical arguments based on the the Higgs mass stability 
tell us that these should be of few hundred GeV, or may be 
few GeV. On the other hand, since masses of particles and 
sparticles have principally different origin, there is no 
physical reason that the particle-sparticle couplings to neutral 
gauginos should be diagonal in a basis of mass eigenstates. 
Thus in general there should be dramatic supersymmetric 
contributions to rates of the flavour changing (FC) processes 
like $\mu\to e+\ga$ decay,  $K^0-\bar K^0$ transition, etc., 
much exceeding the rates predicted by the standard model 
in agreement with experiment. 
Therefore, suppression of such FC transitions in experiment 
puts very strong constraints on the mass and mixing spectrum  
of yet undiscovered sparticles. 

Although in the MSSM natural suppression of the SUSY FC phenomena 
can be achieved  by assuming the universal soft SUSY breaking 
\cite{BFS}, in the context of the LOVE this idea becomes insufficient 
since the effects of physics beyond the GUT scale can strongly violate 
the soft-terms universality at low scales \cite{HKR,BH}.  
These has most dramatic impact on the realistic LOVE frameworks 
like \cite{so10,su6}. 
Such theories above the GUT scale $M_X\simeq 10^{16} GeV$ contain, 
besides the standard chiral set of the `would be light' fermions 
($f=q(u,d);l(\nu,e);u^c,d^c,e^c$), also a vector-like set of 
$F$-fermions $F+\bar F$ in representations similar to that of $f$'s. 
All these states $f,F,\bar F$ interact with constants 
typically $\sim 1$ with various Higgs superfields 
which break GUT symmetry down to $SU(3)\times SU(2)\times U(1)$ 
and also induce large mass terms 
between various fermion states. Actual identification of the light 
physical states $f'$ of the MSSM occurs after integrating out the heavy 
sector \cite{FN} at the GUT scale, and $f'$ generically are 
some linear combinations of the original $f$ and $F$ states. 
These feature provides appealing explanation of the origin of 
fermion flavour, since the small Yukawa constants in MSSM can be 
understood as ratios of different physical scales present in 
the theory beyond the electroweak scale. 
However, the same feature creates severe problems 
is sfermion sector, due to presence of the large Yukawa couplings 
above the GUT scale which generically should not be diagonal 
in the basis of the physical low energy eigenstates $f'$ and 
thus induce the flavour violation. 
The FC effects in the sparticle sector do not decouple: 
even if the SSB terms Plansk scale universality is assumed, 
due to the renormalization group (RG) running effects down from 
$M_{Pl}$ different sfermions will have different (and nondiagonal) soft 
masses at the (GUT) scale where the heavy $F$-states 
decouple \cite{DP}. 
As a result, at lower energies the physical sfermions $\tilde{f}'$ 
will arrive being already strongly split between different 
families, and disoriented relative to fermion $f'$ basis. 
This would give rise to the dramatic FC effects unless the 
SUSY breaking scale is much above the TeV scale. The latter 
case, however, would cancel the advantage of supersymmetry in 
stabylizing the Higgs boson mass. 
Therefore, the SUSY flavour-changing problem poses a serious 
challenge for intimacy of the two (\ref{Love}). 


A natural way to approach (both) flavour problems is to 
consider theories with horizontal (inter-family) symmetry.  
The chiral $SU(3)_H$ symmetry unifying all fermions  
in horizontal triplets \cite{su3H} seems to be most attractive 
for understanding the family replication.  
Here we show that it has a great potential to provide a 
coherent picture of the particle and sparticle masses and 
naturally solve the SUSY FC problem in GUTs. 
(Several other possibilities based on discrete, abelian or 
$U(2)$ family symmetries have been also considered in the 
literature \cite{NS,PT}.) 
For this purpose we revisit a model based on the 
$SU(5)\times SU(3)_H$ symmetry which was suggested in ref. \cite{PLB85}. 
The  model contains $f$ fermions in representations 
$(\bar5+10,3)$: 
\be{f}
\bar{5}_\al=(d^c, l)_\al, ~~~~~ 10_\al=(u^c, q, e^c)_\al 
\ee 
($\al=1,2,3$ is $SU(3)_H$ index). 
Since the fermion mass terms transform as $3\times 3=6+\bar3$, 
they cannot have renormalizable Yukawa couplings to Higgses 
$H=(T,H_2)$, $\bar H=(\bar T,H_1)$ in $(5+\bar5,1)$ representations, 
where $H_{1,2}$ are the MSSM Higgs doublets and $\bar T, T$ 
are their colour-triplet partners.  
Therefore, fermion masses can be induced only via 
higher order operators 
\cite{PLB85}
\be{HOPs}
\frac{g_n\chi_n^{\al\beta}}{M} 10_\al 10_\beta H , ~~~~
\frac{g'_n\chi_n^{\al\beta}}{M}  10_\al \bar 5_\beta \bar H, 
\ee 
where $M\gg M_W$ is some cutoff scale 
(hereafter to be referred as a flavour scale) 
and $\chi_n^{\al\beta}$ denotes a set of `horizontal' Higgs 
superfields in two-index symmetric or antisymmetric representations
of $SU(3)_H$:   
(anti)sextets $\chi^{\{\al\beta\}}$ and triplets 
$\chi^{[\al\beta]}\sim \eps^{\al\beta\ga}\chi_\ga$, 
with VEVs $\langle \chi_n \rangle \leq M$. 
These operators can be effectively induced via exchanges of 
the heavy (with mass $\sim M$) $F$-fermions 
in representations $(10 + \bar 5,\bar 3)$ and $(\ov{10}+5,3)$:  
\beqn{F} 
&{\rm X}^\al  =  (U^c,Q,E^c)^\al, ~~~~
&\bar{\rm X}_\al =(U,Q^c,E)_\al   \nonumber \\
&\bar{\rm V}^\al  =  (D^c,L)^\al ,   ~~~~~ 
& {\rm V}_\al = (D,L^c)_\al 
\eeqn 
(following ref. \cite{PLB85}, we use roman numerals 
${\rm V}=5$ and ${\rm X}=10$ to denote their dimensions).  
Notice, that actual global chiral symmetry of terms (\ref{HOPs}) 
is $U(3)_H=SU(3)_H\times U(1)_H$, where $U(1)_H$ is related to 
the phase transformations $\bar5,10\to e^{i\om}\bar5,10$;  
$\chi_n\to e^{-2i\om}\chi_n$. $U(1)_H$ can serve as 
Peccei-Quinn symmetry unless it is explicitly 
broken in the potential of $\chi_n$ \cite{PLB85}. 

This picture suggests that observed mass hierarchy may emerge 
from the hierarchy of the $U(3)_H$ symmetry breaking VEVs, 
while the form of these VEVs can provide certain predictive 
textures for fermion mass matrices. 
For example, if horizontal Higgses $\chi_n$ are chosen as 
the sextet $\chi$ with a VEV towards (3,3) component, 
and two triplets $\eta$ and $\xi$ having VEVs respectively 
towards 1$^{st}$ and 3$^{rd}$ components \cite{PLB85}: 
\be{VEVs} 
\langle \chi^{\al\beta} \rangle =C \delta^\al_3 \delta^\beta_3, 
~~ \langle \eta_{\ga} \rangle = B \delta_\ga^1, ~~   
\langle \xi_{\ga} \rangle = A \delta_\ga^3 
\ee 
then matrix of their VEVs in whole has a form 
\be{H-VEV} 
\hat{V}_H=\sum_n \langle \chi_n \rangle = 
\matr{0}{A}{0} {-A}{0}{B} {0}{-B}{C} 
\ee
Being projected on the fermion mass pattern via operators (\ref{HOPs}),
this pattern leads to the Fritzsch texture \cite{Fr}. 
(Although the Fritzsch ansatz for fermion masses is already 
excluded by experiment, its viable modification will be 
presented in sect. 3.)  
The inter-family mass hierarchy can emerge from the VEV hierarchy 
$C\gg B\gg A$ in breaking the chiral $U(3)_H$ symmetry
\be{chain}
U(3)_H \stackrel{\chi}{\rightarrow}
U(2)_H \stackrel{\eta}{\rightarrow} 
U(1)_H \stackrel{\xi}{\rightarrow} I
\ee 
rather than due to {\em ad hoc} choice of small Yukawa constants.  

Obviously, the horizontal symmetry can be a good tool also 
for understanding the sfermion mass pattern. In general, it would 
provide the inter-family $SU(3)_H$ degeneracy of the SSB mass 
terms at the flavour scale $M$. However, if the $F$-fermions  
contain all states in (\ref{F}), then inter-family splitting 
between masses of physical sfermions $f'$ will occur after 
integrating out the heavy sector (though with the relatively small 
magnitudes of splittings given by the fermion mass ratios 
between different families), and the sfermion mass matrices 
will be generically disoriented from the fermion mass matrices 
(by angles of the order of the CKM angles). However, even 
under such a "softening" of the problem, the FC effects 
will be persistent and dangerous unless the sfermion masses 
are of several TeV. 

On the other hand, one can observe that the $\wx + \bx$ 
fermions contain all fragments which are needed for generation 
of all quark and lepton masses. Therefore, in principle 
the 5-plets $\bv + \wv$ are not necessary. Below we show 
that if the latter are not introduced (or decoupled at very 
high scales), then fermion and sfermion mass matrices 
induced after integrating out the $\wx + \bx$ states exhibit 
remarkable correlations, and the SUSY flavour-changing 
contributions are naturally suppressed. 


\vspace{5mm} 
\section{Fermion and sfermion masses}  

Let us 
consider a theory which contains the $F$-states only in 
$\wx + \bx$ (\ref{F}). 
The most general renormalizable Yukawa terms have a form:  
\be{Rterms}
W=g10\wx H + f\bar5 \wx \bar H + \wx\Sigma\bx + \bx\chi_n 10 
\ee
(coupling constants $\sim 1$ are understood also in the 
last two terms), 
where $\chi_n$ denotes a set of horizontal Higgs superfields 
transforming as $(R,r)$ representations of $SU(5)\times SU(3)_H$, 
with $R\subset 10\times \ov{10}=1+24+75$ and 
$r\subset \bar3 \times \bar3=3+\bar6$. 
Let us take for simplicity $\Sigma$ as a singlet, though in 
principle it can be in any $(R,r)$ representation 
with $R=1,24,75$; $r=1,8$. Let us also assume that 
$\langle \chi_n \rangle < \langle \Sigma \rangle = M\sim M_X$.

After substituting these VEVs in (\ref{Rterms}), 
the superpotential reduces to the field-dependent mass matrices  
for charged leptons and down quarks: 
\be{Ml}
\begin{array}{cc}
 & {\begin{array}{cc} \,e^c & \,\,\,\;E^c \end{array}}\\ \vspace{2mm}
\begin{array}{c}
l \\ E \end{array}\!\!\!\!\!&{\left(\begin{array}{cc}
0 & \hat{f}H_1 \\ \hat{M}_e^T  & \hat{M} \end{array}\right)} 
\end{array} \!\!\! , ~~~
\begin{array}{cc}
 & {\begin{array}{cc} \,d^c & \,\,\,\;Q^c \end{array}}\\ \vspace{2mm}
\begin{array}{c}
q \\ Q \end{array}\!\!\!\!\!&{\left(\begin{array}{cc}
0 &  \hat{M}_q \\ \hat{f}H_1 & \hat{M} \end{array}\right)} 
\end{array}  
\ee 
and for upper quarks:  
\be{Mu}
\begin{array}{ccc}
 & {\begin{array}{ccc} \,u^c & \,\,\,\;U^c & \,\,\;Q^c 
\end{array}}\\ \vspace{2mm}
\begin{array}{c}
q \\ U \\ Q   \end{array}\!\!\!\!\!&{\left(\begin{array}{ccc}
0 & \hat{g}H_2 & \hat{M}_q \\ 
\hat{M}_u^T  & \hat{M} & 0 \\ 
\hat{g}H_2 & 0 & \hat{M}  \end{array}\right)} 
\end{array}  
\ee 
where each entry is $3\times 3$ matrix in itself. 
As long as $\hat{g}=g\unity$, $\hat{f}=f\unity$ and $\hat{M}=M\unity$  
are flavour blind (unit) matrices, all the
information on the fermion mass and mixing pattern is contained
in the off-diagonal blocks 
$\hat{M}_{q,u,e}=\sum \zeta^n_{q,u,e} \langle \chi_n \rangle$,  
%
where the Clebsch factors $\zeta^n$ depend on the $SU(5)$ content 
of the fields $\chi_n(R,r)$: 
\beqn{zeta}
&&R=1: ~~~~ \zeta_q=\zeta_u=\zeta_e, \nonumber \\ 
&&R=24: ~~~ \zeta_q=-\frac16 \zeta_e, ~~~ \zeta_u=-\frac23 \zeta_e,  
\nonumber \\ 
&&R=75: ~~~ \zeta_q=-\frac13 \zeta_e, ~~~ \zeta_u=\frac13 \zeta_e
\eeqn 
Here we do not impose any constraint on the horizontal VEV pattern, 
assuming only that matrices $\hat{M}_{q,u,e}$ have the hierarchial 
structure resembling that of the fermion mass matrices. 
Implications of the `Fritzsch' texture for the horizontal VEVs 
will be considered in next section. 
 

After integrating out the heavy states theory reduces to 
the MSSM with the light fermion superfields $f'$ 
($f'=q',l',u_c',d_c',e_c'$) which are linear combinations of 
the initial $f$ and $F$ states. 
Notice, that $d^c$ and $l$ (members of $\bar5$) 
do not mix heavy states,
due to absence of $\bv+\wv$ states. 
As for the components of 10: $q,u^c,e^c$, they mix the 
corresponding $F$-states in $\wx$ and in  `seesaw' limit 
$\hat{M}_{q,u,e}\ll M$ 
the states which remain light are 
$q'\simeq q - \hat{M}_q\hat{M}^{-1}Q$, etc. 
 This yields the 
following form of the MSSM Yukawa constant matrices 
below the flavour scale $M$ \cite{PLB85}: 
\beqn{Yukawas} 
&& \hat{\la}^d = f \hat{M}_q \hat{M}^{-1}, ~~~~ 
\hat{\la}^e = f \hat{M}^{-1} \hat{M}_e^T 
\nonumber \\ 
&& \hat{\la}^u = g (\hat{M}_q \hat{M}^{-1} + \hat{M}^{-1} \hat{M}_u^T) 
\eeqn 


Consider now the SSB terms, we do not assume that they 
are universal at the flavour scale $M$. Thus, in general 
trilinear SSB terms have a form repeating all structures present 
in superpotential:  
\be{tri-F}
[A10\wx H + A'\bar5 \wx \bar H + A''\wx\Sigma\bx + A_n \bx\chi_n 10]_F 
\ee
but {\em dimensional} coefficients $A,A'$ etc. are not 
proportional to the Yukawa constants in (\ref{Rterms}).
After substituting the horizontal VEVs, these will reduce 
to terms analogous to (\ref{Ml}) and (\ref{Mu}), but with 
respectively modified entries. 
After integrating out the heavy $F$-sector, we see that trilinear 
SSB terms for the $f'$ states like $\tq \hat{A}_u \tuc H_1$, etc. 
(hereafter we omit $'$ for the light (MSSM) states) 
are alligned to the corresponding Yukawa matrices: 
\be{tri-const} 
\hat{A}_u = \frac{A}{g}\hat{\la}_u, ~~~ 
\hat{A}_d = \frac{A'}{f} \hat{\la}_d, ~~~ 
\hat{A}_e = \frac{A'}{f}\hat{\la}_e 
\ee 


The SSB mass terms of the $f$ and $F$ states at the scale $M$ 
are not universal moreover, however they are degenerated between 
families due to horizontal $SU(3)_H$ symmetry:  
\beqn{soft} 
&m_{\bar5}^2 (\tdc^+\tdc + \tl^+\tl) +  
m_{10}^2 (\tuc^+\tuc + \tec^+\tec + \tq^+\tq)  \nonumber \\ 
&+ m_{\wx}^2 (\TUC^+\TUC + \TEC^+\TEC + \TQ^+\TQ) + \dots
\eeqn 
(if $M<M_X$, the soft masses actually will not be $SU(5)$-invariant,  
but this is not relevant for our discussion.) 
%
After integration out the heavy sector, soft mass terms of  
the physical $f'$ states will get family dependent contributions 
due to mixing between the $f$ and $F$ states.  
%
Taking into account eq. (\ref{Yukawas}), we obtain that 
the sfermion mass matrices have the following form: 
\beqn{softq}
&{\cal M}^2_{\tec}=m_{10}^2\left(\unity + 
\frac{\delta}{f^2} \hat{\la}_e^+\hat{\la}_e\right) , ~~~~~
{\cal M}^2_{\tl,\tdc} = m_{\bar5}^2\unity  \nonumber \\
&{\cal M}^2_{\tuc} = m_{10}^2\left(\unity + 
\delta \hat{\la}^+\hat{\la}\right), ~~~ 
\hat{\la}=\frac{\hat{\la}_u^T}{g} - \frac{\hat{\la}_d^T}{f}
\nonumber \\ 
&{\cal M}^2_{\tq} = m_{10}^2\left(\unity + 
\frac{\delta}{f^2} \hat{\la}_d^+\hat{\la}_d\right)  
\eeqn
where $\delta= (m_{X}^2 - m_{10}^2)/m_{10}^{2}$ is generally $\sim 1$.  

As we see, mass terms of the right down squarks $\tdc$ 
and left sleptons $\tl$ remain degenerate between families 
at the flavour scale $M$. The reason is that the $d^c$ and $l$ 
fragments do not mix $F$-states at the decoupling of the heavy sector 
and thus their soft mass terms maintain the $SU(3)_H$ symmetry. 
Already this fact would be enough to suppress the dominant 
supersymmetric contributions to the $K^0-\bar K^0$ transition 
(gluino-mediated boxes involving $\tdc$ and $\tilde{d}$ states) 
and $\mu \to e \ga$ decay (loops involving $\tl$ states). 

However, in fact flavour conservation appears to be even more 
persistent. Eq. (\ref{softq}) shows that the left down squarks 
$\tilde{d}\subset q$ and right sleptons $\tec$ are split between 
families, but their mass matrices are alligned to the mass matrices 
$\hat{\la}_d$ and $\hat{\la}_e$ of the corresponding fermion states.  
Hence, no flavour changing presents in these sectors. 

As for the upper squarks $\tuc$ and $\tilde{u}\subset q$, 
they are not alligned with $\hat{\la}_u$. Therefore,  
the FC effects will emerge which contribute 
$D^0-\bar{D}^0$ transition, etc. However, the pattern given by 
eq. (\ref{softq}) for upper squarks splitting and mixing is 
by no means in contradiction with the current experimental limits. 
In fact, 
${\cal M}^2_{\tq}$ is alligned to $\hat{\la}_d$, 
so that the mixing angles in $u$-$\tilde{u}$-gluino couplings will 
coincide with the CKM angles. The $u^c$-$\tuc$-gluino
mixing is generally more complicated. Interestingly, 
in the concrete model suggested in sect. 3, 
the right up squarks $\tuc$ appear to be degenerated between 
the first and second families.  
This will lead to further suppression of the FC effects 
in the up quark sector. 

Thus, in the presented framework all dengerous FC 
effects are properly suppressed. However, in severe reality 
these could be induced by the following: 

(i) If $M>M_X$, then owing to the large top Yukawa coupling the 
lepton flavour violation $\mu\to e\ga$, $\tau\to \mu\ga$, 
etc. can be induced due to colour triplet $T$ contribution 
to the RG running from $M$ down to $M_X$ \cite{BH}. 
However, this effect is relevant only for $\tec$ states of 
the third family and presently do not pose any problem.  

(ii) some FC effects could emerge if the seesaw limit is not 
good  in decoupling of the heavy $F$-states 
(see in the concrete model presented in next section). 
However, the corrections to seesaw approximation 
are relevant only for the third family and thus 
should not cause severe problems. 

(iii) MSSM contributions related to the radiative corrections 
below the flavour scale $M$ down to the electroweak 
scale. 
Although in our case 
the boundary conditions for sfermion mass pattern are different 
from the universal boundary conditions generaly adopted in MSSM, 
it is clear that these effects are still under controll. 

(iv) 
FC can be induced by truly non-renormalizable SUSY breaking 
F- and D- terms cutoff by $M_{Pl}$, or the similar terms 
which could be induced after integrating out some other 
$F$-states with mass $\La > M$. However, these effects can be 
suppressed at the needed degree by assuming that $M\ll M_{Pl}$ 
(or $\La$).

Let us conclude this section with the following remark. In the 
above we assumed that some of the $\chi$ fields are in the 
mixed representations of the $SU(5)\times SU(3)_H$.  
For example, in the case of the fields $\chi_n$ having VEVs pattern 
(\ref{H-VEV}) it is natural to take the sextet $\chi$ as $SU(5)$ 
singlet: this will lead to the $b-\tau$ Yukawa unification 
at the GUT scale. However, 
$\eta$ and $\xi$ should be taken in representations 
(24,3) or (75,3), in order to produce the nontrivial Clebsch factors 
which would distinguish the corresponding mass entries between 
quarks and leptons.
However, it would be more economic to think that actually all 
horizontal scalars $\chi,\eta,\xi$ are the $SU(5)$ singlets, 
and the non-trivial Clebsch factors emerge due to the 
higher order operators like $\frac{1}{\La}10\Phi\chi \bx$, 
where $\Phi$ is say standard 24-plet superfield of $SU(5)$.
Clearly, such operators can effectively emerge after integrating 
out some other heavy $F$-states in representations $10+\ov{10}$ 
with masses $\La\gg M,M_X$. In general, this can violate 
the inter-family universality of the soft mass terms at the scale M. 
But not necessarily.  
In particular, there are two possibilities to 
introduce such states and their interactions: 
\beqn{F'}
(A)~  \wx'_\al + \bx'^\al: ~~~~~
10 \Phi \bx' + \La\bx'\wx' + \wx'\chi\bx    \nonumber \\ 
(B)~  \wx'^\al + \bx'_\al: ~~~~~
10 \chi \bx' + \La \bx'\wx' + \wx'\Phi\bx  
\eeqn 
Clearly, in the case (B) the sfermion states in 10 can get 
relatively large non-universal contributions, since the 
different families in 10 mix the states in $\wx'$ with 
different angles.  This could induce the considerable 
flavour violation. However, 
in the case $(A)$ mixing between the states in 10 and $\wx'$ 
is universal in families, at least in leading approximation. 
Therefore, the mass terms in (\ref{soft}) will not receive 
large family dependent contributions and remain degenerate.  
 Concluding, although the group-theoretical structure of the 
considered effective operator does not depend on the 
pattern and way of exchange of $F$-states, the soft mass 
terms pattern is sensitive with respect to the latter.

\vspace{5mm}
\section{An example of consistent and predictive model} 

In the framework presented above the solution of the flavour 
changing problem practically does not depend on the concrete 
structure of the fermion mass matrices (i.e. $SU(3)_H$ breaking 
VEVs pattern). However, one can already notice that considered  
general framework is capable to produce the predictive 
ansatzes for fermion masses, which capability is rather resembling 
that of the $SO(10)$ model. Indeed, since the mass matrices 
of {\em all} quark and lepton masses emerge from the {\em same} 
$F$-fermion exchange, they will actually differ only by the 
Clebsch structure of the "vertical" Higgss VEVs responsible for 
the GUT symmetry breaking - as it in fact takes place in $SO(10)$ 
model.  

Here we would like to present an example of consistent and 
predictive model built along the lines discussed above. 
The requirements to be fulfilled are: 

(i) Natural doublet-triplet splitting 

(ii) Predictivity in fermion mass and mixing pattern 

(iii) Predictivity in sfermion mass and mixing and natural 
suppression of the SUSY FC contributions 

(iv) Natural suppression of the proton decaying $d=5$ operators 
(which in certain sence are the part of the whole flavour changing 
problem).  

Let us start from the first requirement. In order to achieve 
the solution to the doublet-triplet splitting problem, 
we extend the vertical symmetry to $SU(5)\times SU(5)'$ 
\cite{Maslik}. The model involves the Higgs superfields 
$\OM_a(\bar5,5,1)$ and $\bar{\OM}_a(5,\bar5,1)$ ($a=1,..4$), 
with VEVs of the following possible structures: 
\beqn{OM}
&& \langle\OM_1\rangle=M_1\cdot {\rm diag}(1,1,1,1,1),  \nonumber \\ 
&& \langle\OM_2\rangle=M_2\cdot {\rm diag}(0,0,0,1,1),  \nonumber \\ 
&& \langle\OM_3\rangle=M_3\cdot {\rm diag}(1,1,1,0,0),  \nonumber \\ 
&& \langle\OM_4\rangle=M_4\cdot {\rm diag}(y,y,y,1,1), ~~~ y\neq 1 
\eeqn  
In order to maintain the gauge coupling unification, it is 
natural to assume that $M_1 \gg M_{2,3,4}\sim M_X$. In this case 
$SU(5)\times SU(5)'$  at the scale $M_1$ first reduces 
to the diagonal $SU(5)$ subgroup, which then breaks down 
to $SU(3)\times SU(2)\times U(1)$ at the scale $M_X$. 
Let us also introduce Higgses in representations 
$H(5,1) + \bar H(\bar5,1)$ and $H'(1,5) + \bar H'(1,\bar5)$. 
Imagine now that these are coupled to the $\OM$ fields through 
the following terms in superpotential: 
\be{DT} 
\bar H \OM_{1(4)} H'  +  H \bar{\OM}_3 \bar H'
\ee
after substituting the VEVs (\ref{OM}), we see that just two 
MSSM Higgs doublets remain light: $H_2 \subset H$ and 
$H_1 \subset \bar{H}'$, while all colour triplets and other couple 
of doublets get masses $\sim M_X$. The doublet-triplet splitting 
is achieved in this way. 

Let us introduce now the horizontal $SU(3)_H$ symmetry. 
We assume that the fermion superfields $\bar5_\al$ and 
$10_\al$ (\ref{f}) belong to representations $(\bar5+10,1,3)$.  
The Horizontal Higgses are taken as a sextet $\chi$ and two 
triplets $\eta,\xi$, with the VEVs $\sim M_X$. We assume that 
their VEVs  have the structure (\ref{H-VEV}), which can be indeed 
obtained in analysis of their superpotential \cite{BDJC}. 
$\Sigma$ is taken as an octet of $SU(3)_H$ 
with a VEV towards the $\la_8$ generator \cite{AB}: 
$\langle \Sigma \rangle = M{\rm diag}(1,1,-2)$,  
$M\geq A,B,C$.  
All these are singlets of $SU(5)\times SU(5)'$. 

We assume further that the only fermion states present 
at the scale $M$ besides the chiral $f$ fermions $\bar5+10$, 
are the states $\wx^\al + \bx_\al$, which acquire mass via 
$\Sigma$, and by some symmetry reason only the Higgses 
$\chi$ and $H$ couple these fermions in a renormalizable way: 
\be{Rterms1}
10\chi\bx +   \bx \Sigma \wx + g \wx 10 H 
\ee 
while other Higgses interact via higher order operators: 
\be{NRterms} 
10 \frac{\OM_2 \bar{\OM}_2}{\La^2} \eta \bx + 
10 \frac{\OM_4 \OM_4 \bar{\OM}_4 \bar{\OM}_4 }{\La^4} \xi \bx + 
g'\wx \bar5 \frac{\OM_2}{\La} \bar{H}'  
\ee 
where $\La \gg M$ is some cutoff scale, say $\La\sim M_1$. 
Needless to say that these operators can be effectively induced 
via integrating out the $F$-states with masses $\sim \La$ along 
the lines discussed at the end of previous section, 
so that they do not introduce the flavour violating effects. 

 Then, by taking into account the VEV pattern (\ref{H-VEV}), 
from eqs. (\ref{Yukawas}) in the seesaw approximation 
we arrive to the following textures: 
\beqn{Fr-new}
&&\hat{\la}^u=\frac{g}{M} 
\matr{0}{\kappa_uA}{0}{-\kappa_uA}{0}{\frac12\eps_uB}{0}{\eps_uB}{C} 
\nonumber \\ 
&&\hat{\la}^d=\frac{\eps g'}{M} 
\matr{0}{\kappa_dA}{0}{-\kappa_dA}{0}{\frac12\eps_dB}{0}{\eps_dB}{\frac12C}   
\nonumber \\ 
&&\hat{\la}^e=\frac{\eps g'}{M} 
\matr{0}{-\kappa_eA}{0}{\kappa_eA}{0}{\eps_eB}{0}{\frac12\eps_eB}{\frac12C} 
\eeqn 
where $\eps=M_2/\La$,  and $\eps_{u,d,e}\sim \eps^2$, 
$\kappa_{u,d,e}\sim \eps^4$ contain Clebsch factors emerging from 
the first two term in (\ref{NRterms}). 
In particular, the structure of the VEV $\OM_2$ leads to 
$\eps_u=\eps_d=\frac12 \eps_e$, while $\kappa_{u,d,e}$ in general 
are not related due to complex structure of the $\OM_4$ and 
$\bar{\OM}_4$ VEVs. 

The fact that $H_1$ couples to fermions via higher order operator, 
allows us to assume that the constant $g'$ is order 1 as well as 
the Yukawa constant $g\sim \la_t$, and at the same time 
to have small $\tan\beta$. The smalness of the bottom-tau 
masses with respect to top in this case can be attributed to 
the suppression factor $\eps=M_2/\La$ rather than to large $\tan\beta$. 
On the other hand, the structure of $\langle \OM_2\rangle$ 
implies that colour triplet $\bar{T}'$ in $\bar{H}'$ is decoupled 
from $\bar5$ states and thus proton is perfectly stable in this theory. 
Interesting enough, eq. ({\ref{softq}) indicates that the $\tuc$ 
mass matrix ${\cal M}^2_{\tuc}$ is degenerated between first 
two families, which provides further suppression of the 
flavour changing and CP-violating phenomena in the up sector 
as compared to general case of sect. 2.

The mass matrix texture, which depends on 7 real parameters, 
allows to make predictions. In a leading approximation, 
we obtain the following relations between the MSSM Yukawa 
constants at the GUT scale: 
\be{btau}
\la_b=\la_\tau, ~~~~~
\frac{\la_c}{\la_t} = \frac14 \frac{\la_s}{\la_b} = 
\frac{1}{16} \frac{\la_\mu}{\la_\tau}    
\ee 
and the expressions for the CKM mixing angles 
\beqn{Fr-angles}  
&&s_{12}=\left|\sqrt{\frac{\la_d}{\la_s}}-
e^{i\delta}\sqrt{\frac{\la_u}{\la_c}}\right|, \nonumber \\ 
&&s_{23}=\sqrt{\frac{\la_s}{2\la_b}}-\sqrt{\frac{\la_c}{2\la_t}}=
\frac14 \sqrt{\frac{\la_\mu}{2\la_\tau}} = 0.043 
\nonumber \\ 
&&\frac {s_{13}}{s_{23}} = \left|\frac {V_{ub}}{V_{cb}}\right| =
\sqrt{\frac{\la_u}{\la_c}}\, ~~~~ ~~~~
\left|\frac {V_{td}}{V_{ts}}\right| =\sqrt{\frac{\la_d}{\la_s}}\,
\eeqn 
where $\delta$ is a CP-violating phase. In particular, 
when $\delta\sim 1$, we obtain $s_{12}\approx \sqrt{m_d/m_s}$. 
The excat implications of these GUT scale relations 
for the low energy observables, though almost obvious, 
will be given elsewhere. 

Let us conclude this section with the following remark. 
The seesaw limit is certainly good for first two families. 
However, the fact that $\lambda_t\sim 1$ implies for the 33 
entry of matrices $\hat{\la}_{u,d,e}$ $C/M\sim 1$ 
unless the coupling constant $g$ is much above the perturbativity 
bound. Thus for the Yukawa constants of the third generation  
one has to use the exact formula (see e.g. in \cite{Rattazzi}). 
Then the genuine Yukawa constants of the third family 
$\tilde{\la}_{t,b,\tau}$ are related to the 
`would-be' Yukawa constants $\la_{t,b,\tau}$ given in the 
seesaw limit (\ref{Yukawas}) as 
\be{la-t}
\tilde{\la}_{t,b,\tau}=\frac{\la_{t,b,\tau}}
{\sqrt{1+(\la_t/2g)^2}}<\la_{t,b,\tau}
\ee 
Therefore, for $\la_t\simeq 1$, as it is favoured from eqs. 
(\ref{btau}), implies that seesaw limit can be reasonably good 
already for $g=1$.

To conclude, it seems that in order to deal with the fermion 
flavour and sflavour problems in a coherent way, 
All you need is LOVE (SUSY+GUT), 
though with a little help of little friend \cite{LM} 
which comes from the horizontally understood symmetry properties.  




\end{document}